\documentclass[11pt]{article}
\usepackage{moriond,epsfig,wrapfig}

\def\Journal#1#2#3#4{{#1} {\bf #2} (#3) #4}

\def\NPB{{\em Nucl. Phys.}   {\bf B}}
\def\PLB{{\em Phys. Lett.}   {\bf B}}

\def\PRD{{\em Phys. Rev.}    {\bf D}}
\def\ZPC{{\em Z. Phys.}      {\bf C}}

\def\CPC{\em Comp. Phys. Commun.}

\newcommand{\etal}{{\em et al.}}

\def\be{\begin{equation}}
\def\ee{\end{equation}}
\def\bea{\begin{eqnarray}}
\def\eea{\end{eqnarray}}

\newcommand{\pom}{{I\!\!P}}

\newcommand{\xpom}{\ensuremath{{x_\pom}}}
\newcommand{\fpom}{\ensuremath{f_{\pom}}}

\newcommand{\apom}{\ensuremath{\alpha_\pom}}
\newcommand{\appom}{\ensuremath{\alpha'_\pom}}
\newcommand{\ftd}{\ensuremath{{F_2^D}}}
\newcommand{\fld}{\ensuremath{{F_L^D}}}
\newcommand{\srd}{\ensuremath{{\sigma_r^D}}}
\renewcommand\figurename{Fig.}

\newcommand{\eqref}[1]{Eq.~(\ref{#1})}
\newcommand{\figref}[1]{\figurename~\ref{#1}}

\newcommand{\etjet}{%
\ensuremath{%
{E_T^{\rm jet}}}}

\newcommand{\xgammajets}{%
\ensuremath{%
{x_\gamma^{\rm jets}}}}
\newcommand{\xgamma}{%
\ensuremath{%
{x_\gamma}}}
\newcommand{\zpomeronjets}{%
\ensuremath{%
{z_\pom^{\rm jets}}}}

\begin{document}
\vspace*{4cm}
\title{MEASUREMENTS OF THE STRUCTURE OF DIFFRACTION AT HERA}

\author{Sebastian Sch\"atzel\\(on behalf of the H1 and ZEUS Collaborations)}

\address{DESY FLC, Notkestr. 85, D-22607 Hamburg, Germany}

\maketitle
%%%%%%%% authors photo (option) %%%%%%%%%%%%%%%%%%%%% 
\vspace{-.2cm}
\begin{center}
\framebox[35mm]{\rule[-11mm]{0mm}{35mm}}
\end{center}

\vspace*{0.2cm}
\abstracts{%
Measurements of diffraction in $ep$ collisions are
presented. 
All measured processes
which occur through the exchange of a virtual photon (DIS)
are consistent with a universal structure.
For the diffractive photoproduction of dijets, factorisation is broken
by a global factor $\approx 0.5$ with respect to the DIS processes.}

\vspace*{-1.0cm}
\section{Inclusive diffraction at HERA}
\begin{wrapfigure}{l}{4.8cm}
\vspace*{-0.4cm}
\includegraphics[width=0.3\textwidth,keepaspectratio]{%
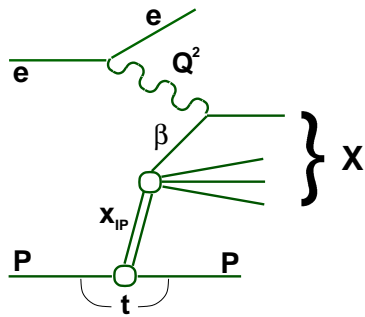}
\vspace*{-0.8cm}
\caption{Diffractive $ep$ scattering.}
\label{fig1}
\vspace*{-0.3cm}
\end{wrapfigure}

\vspace*{-0.15cm}
At the HERA $ep$ collider the diffractive quark structure of the proton
is probed with a point-like photon (\figref{fig1}a). 
The hard scale of the interaction is given by the photon virtuality $Q^2$.
%\begin{wrapfigure}{l}{10.9cm}
%\setlength{\unitlength}{1cm}
%\begin{picture}(16,4.6)
%\put(0,0){%
%\includegraphics[width=0.35\textwidth,keepaspectratio]{%
%fig1.eps}
%}
%\put(5.3,0){%
%\includegraphics[width=0.35\textwidth,keepaspectratio]{%
%fig1cc.eps}
%}
%\put(4,0.3) {\textbf{a)}} 
%\put(9.5,0.3) {\textbf{b)}}
%\end{picture}
%\vspace*{-0.8cm}
%\caption{Diffractive $ep$ scattering. a) Neutral current, b) charged current process.}
%\label{fig1}
%\vspace*{-0.3cm}
%\end{wrapfigure}
%
In diffraction the proton remains intact and
loses only a small fraction $\xpom$ of its initial momentum.
Such events can be selected experimentally by requiring an empty 
area in the detector (rapidity gap) between the outgoing proton 
and the produced hadronic system $X$.
Since the outgoing proton is not detected, the measured cross
section $\sigma^D$ is integrated over $|t|<1$~GeV$^2$, where 
$t$ is the squared 4-momentum transferred at the proton vertex, and
the measured cross sections include
$\approx 10\%$ of events where the proton is excited into a 
system of mass $<1.6$~GeV.
In a two-step picture,  
the proton exchanges a diffractive object (often called
the pomeron) with momentum fraction $\xpom$ and the quark struck by the 
photon carries a fraction $\beta$ of the momentum of the diffractive
exchange. Additional kinematic variables are Bjorken-$x$ $x=\beta\, \xpom$ and
the inelasticity $y=Q^2/(s x)$ where $s$ is the $ep$ centre-of-mass
energy squared.
The reduced diffractive cross section $\srd$ is given by
$\frac{d^3 \sigma^D}{d\xpom\,dx\,dQ^2}=\frac{4\pi\alpha^2}{xQ^4} 
Y\, \srd(\xpom,x,Q^2),$
with $Y \equiv \left( 1-y+\frac{y^2}{2} \right)$, and 
is related to the structure functions by
$\srd=\ftd - \frac{y^2}{2Y} \fld$.
The diffractive structure functions have been proven to factorise
into diffractive quark densities $f_i^D$ of the proton convoluted
with ordinary photon-quark scattering cross sections 
$\hat{\sigma}^{\gamma^* i}$:
$\ftd = \sum_i f_i^D \otimes \hat{\sigma}^{\gamma^* i}$,
where the sum runs over all contributing quark and anti-quark 
flavours.\cite{Collins} 
This leading twist factorisation formula holds
for large enough $Q^2$ and applies also to $\fld$.
Diffractive parton densities (PDFs) which are applicable at fixed
$\xpom$, obey the standard QCD evolution
equations and can be determined from fits to structure function data.

\vspace*{-0.3cm}
\section{Inclusive Measurements and Parton Densities}
\vspace*{-0.15cm}
\begin{figure}
\setlength{\unitlength}{1cm}
\begin{picture}(16,8)
\put(0,0){%
\includegraphics[width=9cm,keepaspectratio]{%
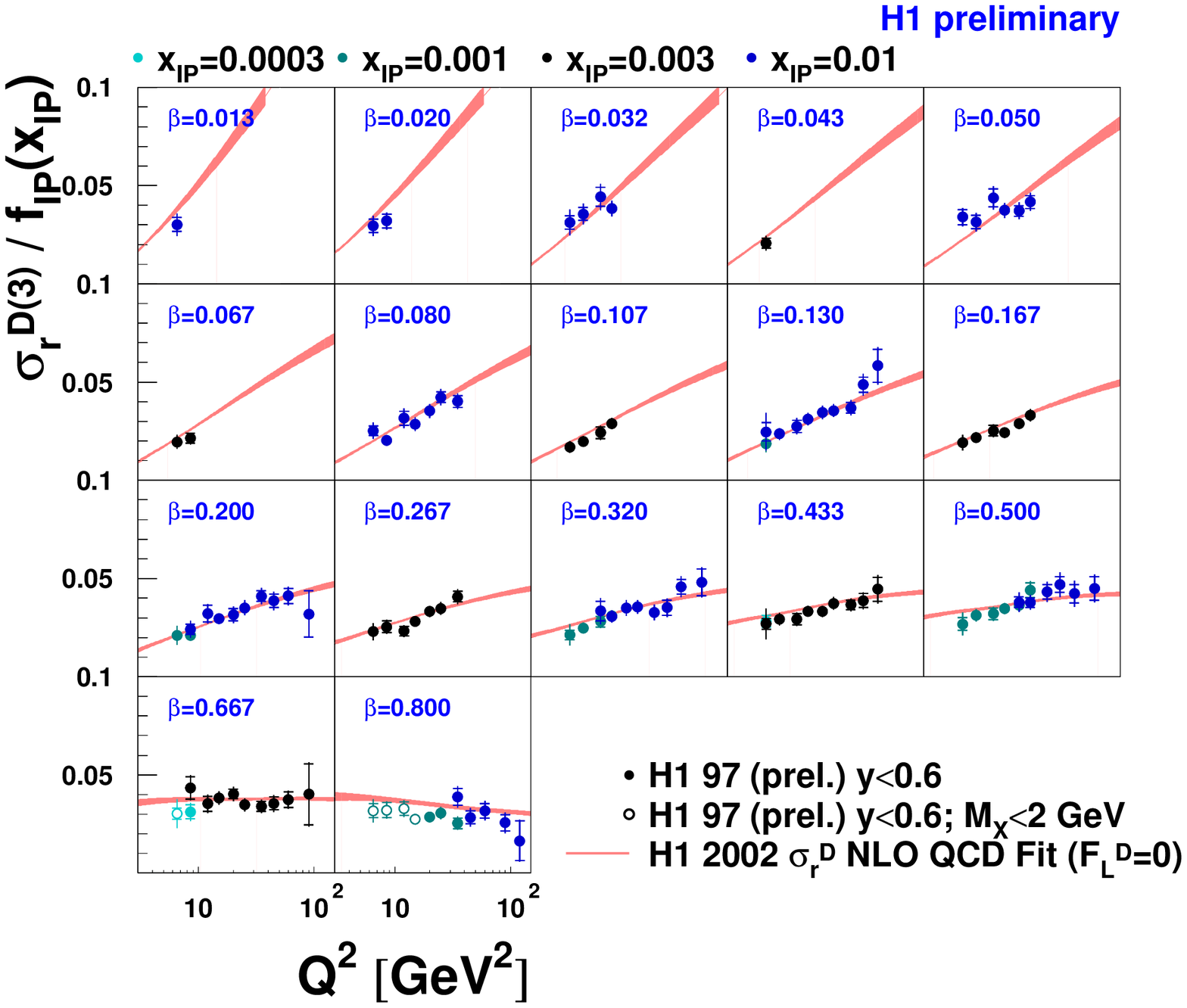}
}
\put(9.,1){%
\includegraphics[width=7cm,keepaspectratio]{%
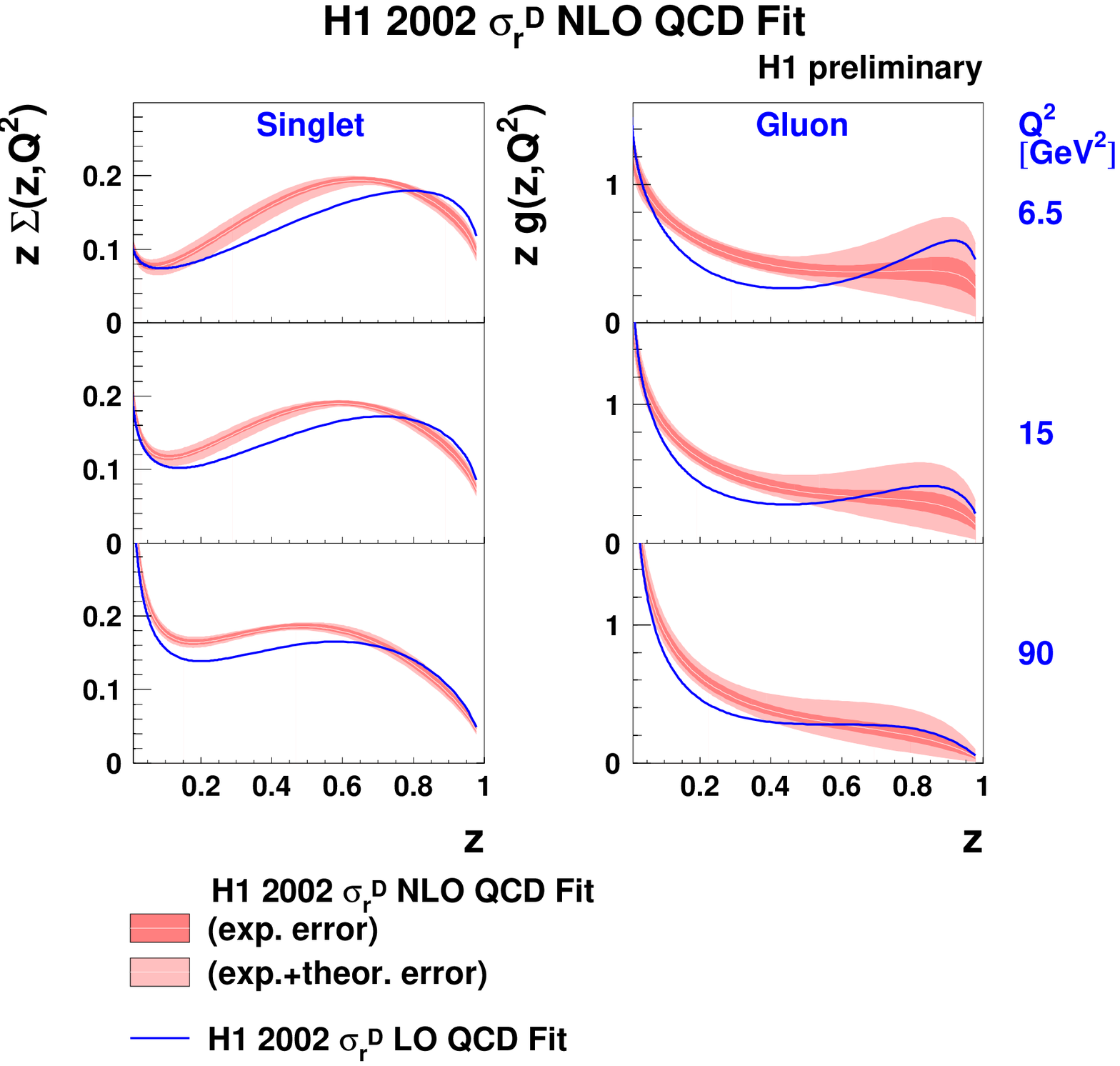}
}
\put(0.5,0.1) {\textbf{a)}} 
\put(9.5,0.1) {\textbf{b)}}
\end{picture}
\vspace*{-0.8cm}
\caption{a) Reduced diffractive cross section divided by the
  flux factor as a function of
  $Q^2$ compared with the H1 2002 NLO fit.
b) Diffractive quark (``Singlet'') and gluon densities.}
\label{figqs}
\vspace*{-0.5cm}
\end{figure}
Diffractive structure function measurements have been performed by the
H1 and ZEUS Colla\-borations.\cite{fit,zeusf2d}
The H1 Collaboration has extracted diffractive PDFs from their data.
The collected event sample statistics do not allow an extraction
at fixed values of $\xpom$.
The $\xpom$ and $t$ dependences of the PDFs are therefore parameterised 
in a flux factor \fpom:
$f_i^D(\beta, Q^2, \xpom, t) = f_{\pom}(\xpom,t) \ f^D_i(\beta, Q^2),$
with
 $\fpom(\xpom,t) = e^{B t} \xpom^{1-2 \apom(t)}$  
where $\apom(t) = \apom(0) + \appom\,t$ is the linear
pomeron Regge trajectory and $\apom(0)=1.17^{+0.07}_{-0.05}$. 
The flux factor approach is consistent with the diffractive data within
uncertainties for $\xpom<0.01$. 
At larger $\xpom$ values, a second term has to be introduced which can be
interpreted as subleading reggeon exchange.
The reduced cross section divided by the flux factor is shown 
in \figref{figqs}a as a function of $Q^2$ for different $\beta$ values
in a kinematic region
where to a good approximation $\srd = \ftd$ 
and the reggeon term is negligible. 
The cross section exhibits scaling violations with positive $\partial
\srd/\partial \ln Q^2 $ up to $\beta \approx
2/3$ which are driven by the gluon distribution. 
Also shown is 
the DGLAP QCD fit and 
\begin{wrapfigure}{r}{6cm}
\vspace*{-0.2cm}
\includegraphics[height=4.7cm,keepaspectratio]{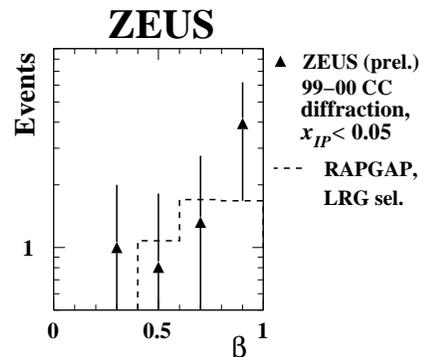}
\vspace*{-0.45cm}
\caption{Diffractive charged current events as a function
  of $\beta$ compared with a LO Monte Carlo prediction.}
\vspace*{-1cm}
\label{figcc}
\end{wrapfigure}
the extracted quark and gluon densities are shown in \figref{figqs}b.
The gluon carries $\approx 75\%$ of the momentum of the diffractive
exchange and is poorly known at large fractional momentum $z$.
Diffractive NLO PDFs have also been obtained by Alvero \etal\,\cite{actw}
in combined fits to H1 and ZEUS diffractive data.

\vspace*{-0.3cm}
\section{Charged current events}
\vspace*{-0.15cm}
Diffractive processes which occur through $W$ boson exchange 
have been measured by ZEUS and H1 using events
with missing transverse energy which is carried away by a neutrino.\cite{zeuscc,cc}
Both experiments find the ratio of the diffractive to the 
inclusive charged current cross section to be $\approx 3\% \pm 1\%$ 
for $\xpom<0.05$ and $Q^2>200$~GeV$^2$.
The $\beta$ distribution of the events as measured by the ZEUS
Collaboration is shown in \figref{figcc}.
It is well described by a leading order Monte Carlo prediction
obtained with the RAPGAP program\,\cite{rapgap}
using an earlier version (``H1 fit 2'') of the diffractive PDFs.\cite{h1fit2} 

\vspace*{-0.3cm}
\section{Final State Measurements}
\begin{wrapfigure}{r}{8cm}
\setlength{\unitlength}{1cm}
%\vspace*{+0.6cm}
\begin{picture}(16,4)
\put(0,0){%
\includegraphics[width=4cm,keepaspectratio]{%
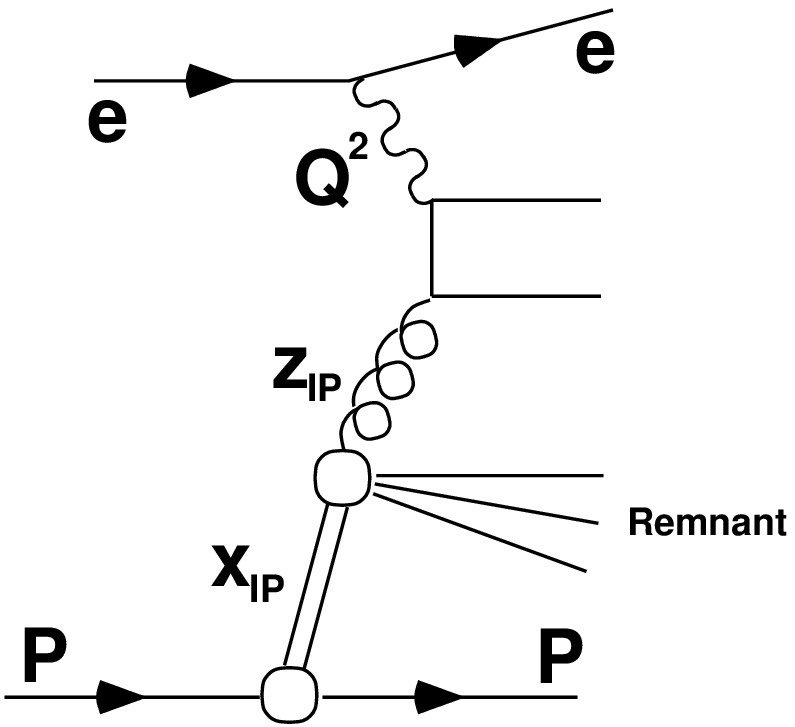}
}
\put(4,0){%
\includegraphics[width=4cm,keepaspectratio]{%
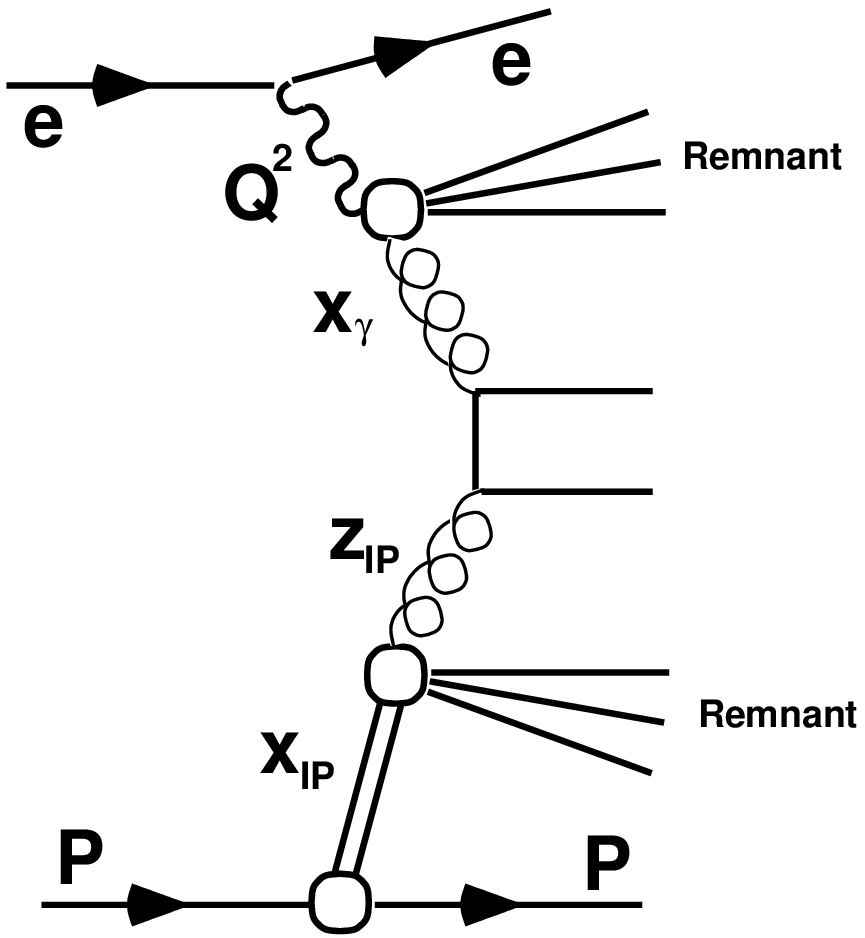}
}
\put(0,2) {\textbf{a)}} 
\put(4,2) {\textbf{b)}}
\end{picture}
\vspace*{-0.8cm}
\caption{Example processes for a) direct photon and b)
 resolved photon interactions.}
\label{fig:bgf}
\vspace*{-0.8cm}
\end{wrapfigure}

\vspace*{-0.15cm}
Diffractive dijet and D$^*$ meson (heavy quark) production are
directly sensitive to the diffractive gluon through the photon-gluon 
fusion production  mechanism (\figref{fig:bgf}a)  and are
 used to test 
the universality of the diffractive parton densities.

\vspace*{-0.3cm}
\subsection{Diffractive dijet production in DIS}
\begin{wrapfigure}{r}{11.3cm}
\setlength{\unitlength}{1cm}
\begin{picture}(16,5)
\put(0,0){%
\includegraphics[width=0.35\textwidth,keepaspectratio]{%
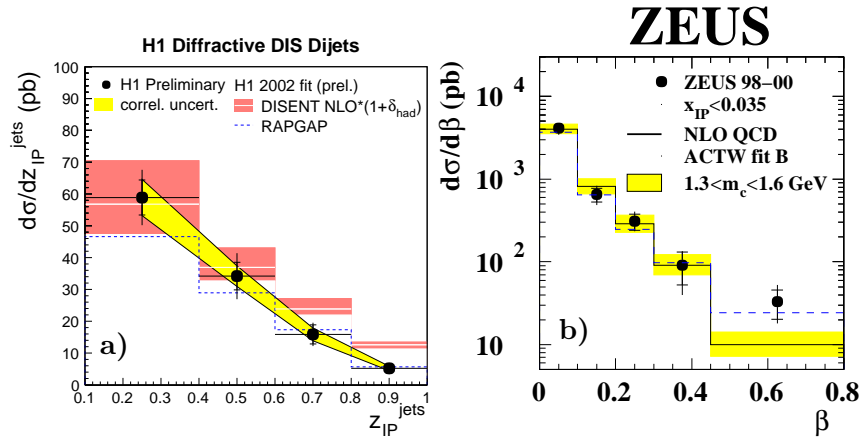}
}
\put(5.7,0){%
\includegraphics[width=0.35\textwidth,keepaspectratio]{%
zeus_dstar_beta.eps}
}
\put(1.2,1.1) {\textbf{a)}} 
\put(7.3,1.3) {\textbf{b)}}
\end{picture}
\vspace*{-0.8cm}
\caption{Diffractive DIS cross section for a) dijet production b)
  $D^*$ production compared with NLO predictions based on diffractive PDFs.}
\label{dis}
\vspace*{-0.2cm}
\end{wrapfigure}

Cross sections for dijet production in the kinematic
range $Q^2>4$~GeV$^2$, $\etjet(1,2)>5,4$~GeV
have been measured by H1\,\cite{disjets} using
the inclusive $k_T$ cluster 
algorithm to identify jets.
NLO predictions have been obtained by interfacing the 
H1 diffractive PDFs with
the \mbox{DISENT} program.\cite{disent}
The renormalisation and factorisation scales were set to the
transverse energy of the leading parton jet.
The NLO parton jet cross sections have been corrected for
hadronisation effects using RAPGAP with parton 
showers
and Lund string 
fragmentation.
Comparisons of the DISENT and RAPGAP predictions with the measured
cross section differential in $\zpomeronjets$, an estimator for the
fraction
of the momentum of the diffractive exchange entering the hard scatter,
 are 
shown in \figref{dis}a.
The inner band around the
NLO calculation indicates the $\approx 20\%$ uncertainty resulting
from a variation of
the renormalisation scale by factors 0.5 and 2. 
The uncertainty in the diffractive PDFs is not shown. Taking
this additional uncertainty into account leads to a large increase in
the uncertainty at high \zpomeronjets, such that the dijet cross section
is well described by the NLO calculation based on the diffractive PDFs.

\vspace*{-0.3cm}
\subsection{Diffractive $D^*$ production in DIS}
The diffractive production of $D^*$ mesons has been
measured by the ZEUS and H1 Collaborations.\,\cite{zeusdstar,dstar}
\figref{dis}b shows the ZEUS cross section ($1.5 < Q^2< 200$~GeV$^2$, $\xpom
 < 0.035$,
$p_T^{D*}>1.5$~GeV/c) as a function of $\beta$
compared with NLO predictions of a diffractive extension of the
HVQDIS program\,\cite{hvqdis} interfaced to 
the Alvero \etal{} fit B diffractive PDFs. 
The charm mass is $m_c=1.45$~GeV, the QCD scales 
are $\mu=\sqrt{Q^2 + 4 m_c^2}$ and
the Peterson fragmentation function was used with
$\varepsilon=0.035$. The NLO error band is given by 
variations of  $m_c$ from 1.3 to 1.6~GeV.
The cross section is well described by the
NLO calculation within errors.
The H1 data on diffractive $D^*$ production is well described 
within errors by the H1 2002 NLO diffractive PDFs.\cite{dstar}

\subsection{Diffractive photoproduction of dijets}
\begin{wrapfigure}{l}{11.3cm}
\setlength{\unitlength}{1cm}
\begin{picture}(16,6.3)
\put(3.3,5.1){%
\includegraphics[width=0.4\textwidth,keepaspectratio]{%
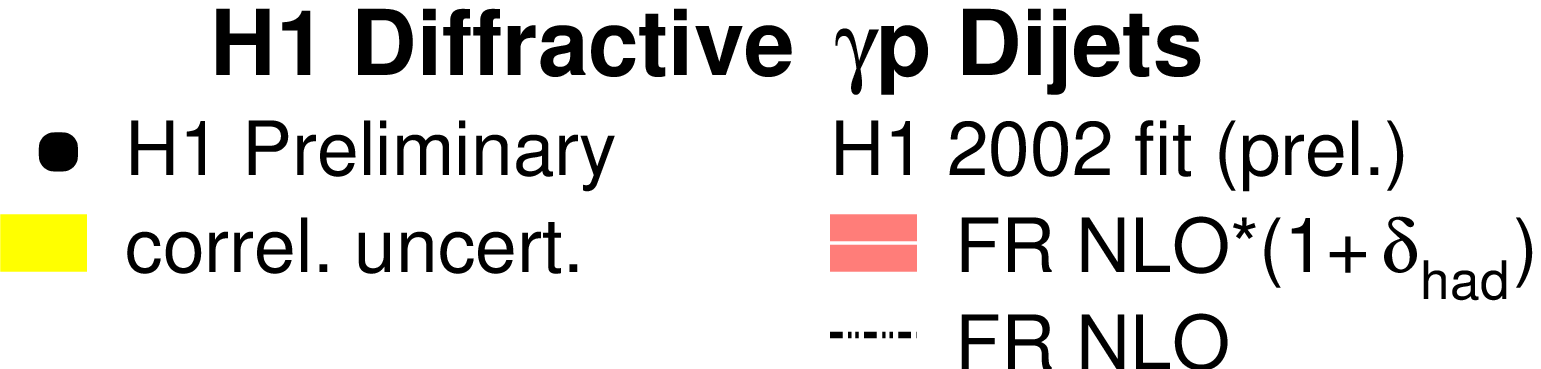}
}
\put(0,0){%
\includegraphics[width=0.35\textwidth,keepaspectratio]{%
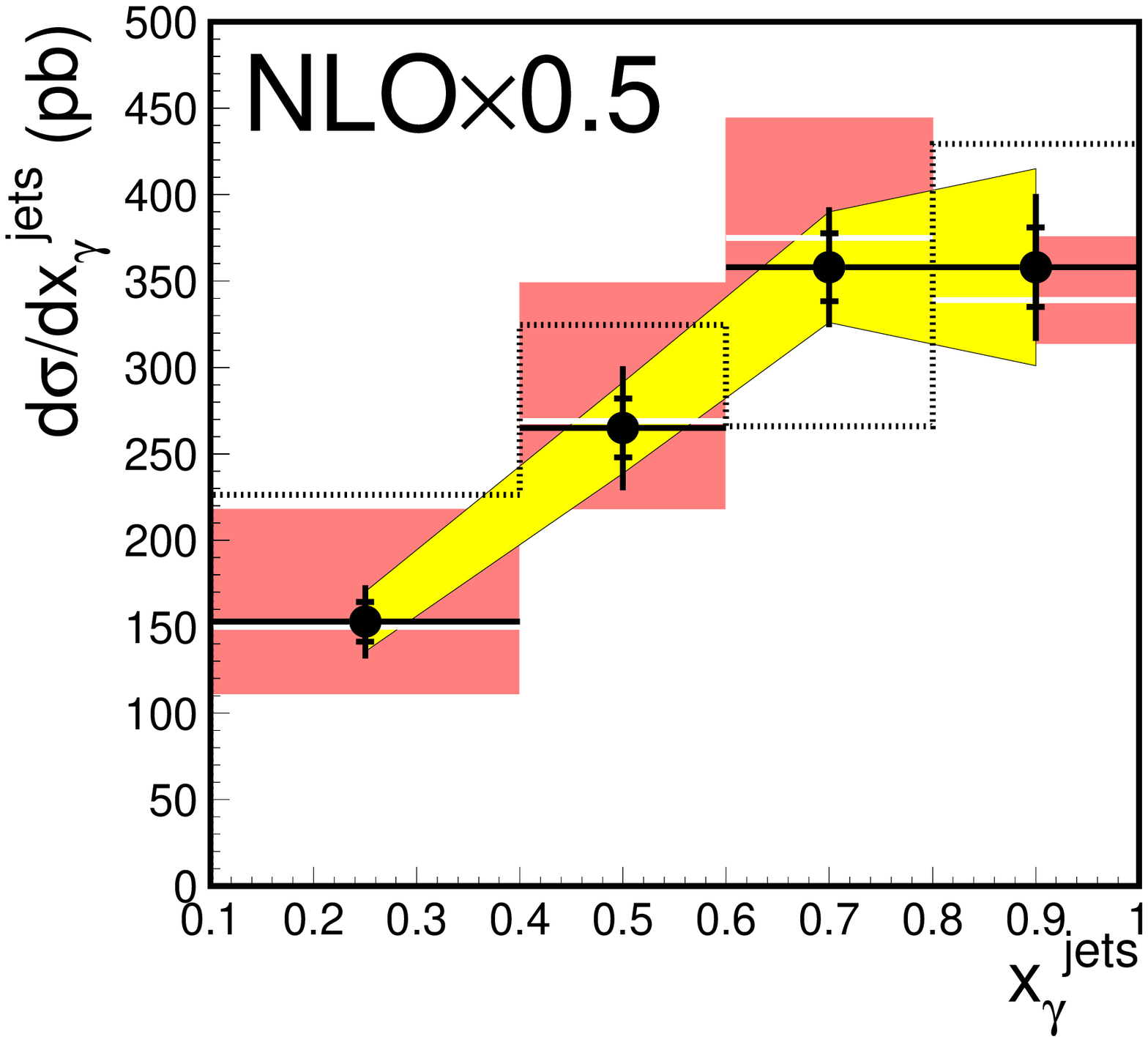}
}
\put(5.7,0){%
\includegraphics[width=0.35\textwidth,keepaspectratio]{%
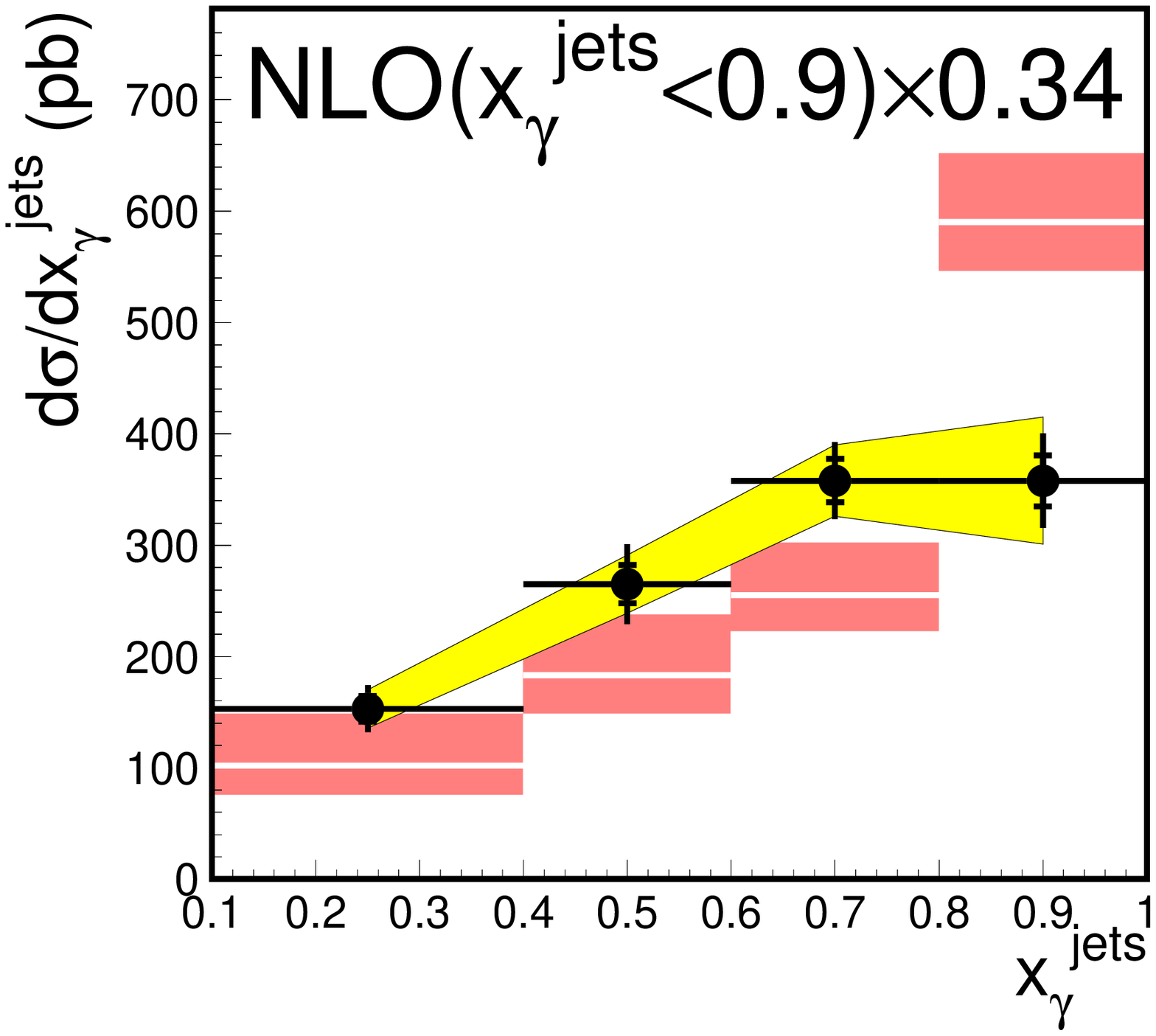}
}
\put(1.2,1.1) {\textbf{a)}} 
\put(10.5,1.1) {\textbf{b)}}
\end{picture}
\vspace*{-0.8cm}
\caption{Diffractive dijet photoproduction cross section.
a) NLO calculation scaled by global factor 0.5, b) ``resolved''
part scaled by 0.34.}
\label{gp}
\vspace*{-0.2cm}
\end{wrapfigure}

Dijet cross sections in diffractive photoproduction ($Q^2 \approx 0$)
have been measured by the H1 and ZEUS Collaborations.\,\cite{disjets,zeusgp}
In photoproduction, a sizeable contribution to the cross section is
given by resolved photon processes
(\figref{fig:bgf}b) in which only a fraction $\xgamma <1$ of the
photon
momentum enters the hard scatter.
The H1 cross section 
($Q^2<0.01$~GeV$^2$, $\xpom<0.03$, $\etjet(1,2)>5,4$~GeV,
inclusive $k_T$ algorithm)
is shown in \figref{gp}. NLO predictions have
 been obtained with the
Frixione \etal{} program\,\cite{frixione} interfaced to the H1 diffractive
PDFs. The parton jet calculation is corrected for hadronisation effects
using RAPGAP.
The calculation lies a factor $\approx 2$ above the data and
factorisation is therefore broken in photoproduction.
\figref{gp}a shows the NLO predictions scaled down by a global factor
0.5 which gives a good description of the measured cross section.
In \figref{gp}b, only the ``resolved'' part for which $\xgammajets <
0.9$ at the parton level is scaled by a factor 0.34, as proposed
by Kaidalov \etal\,\cite{kaidalov} The calculation for
$\xgammajets>0.9$ is left unscaled. This approach is clearly disfavoured.
The same conclusion is reached by the ZEUS analysis in a comparison
with NLO calculations by Klasen and Kramer.

\vspace*{-0.3cm}
\section{Conclusions}
\vspace*{-0.15cm}
Diffractive deep-inelastic $ep$ scattering is described
by universal parton densities. For dijets in diffractive 
photoproduction, factorisation is broken by
a factor $\approx 0.5$ with no observed dependence on $\xgammajets$ or
other kinematic variables.

%\section*{Acknowledgments}


\vspace*{-0.5cm}
\section*{References}
\small
\vspace*{-0.25cm}
\begin{thebibliography}{99}
\bibitem{Collins} 
J.C.~Collins, \Journal{\PRD}{57}{1998}{3051}; erratum ibid. {\bf D61}
(2000) 019902.
\bibitem{fit}
H1 Collaboration, paper 980 subm. to ICHEP 2002.
\bibitem{zeusf2d}
ZEUS Collaboration, \Journal{\NPB}{713}{2005}{3}.
\bibitem{actw} 
L.~Alvero,  J.C.~Collins, J.~Terron, J.J.~Whitmore, \Journal{\PRD}{59}{1999}{074022}.
%\bibitem{dglap} V.~Gribov, L.~Lipatov, \Journal{\SJNP}{15}{1972}{438,
%    675}; \\
%Y.~Dokshitzer, \Journal{\SPJ}{46}{1977}{641};\\
%G.~Altarelli, G.~Parisi, \Journal{\NPB}{126}{1977}{298}.
\bibitem{zeuscc}
ZEUS Collaboration, paper 6-0229 subm. to ICHEP 2004.
\bibitem{cc}
H1 Collaboration, paper 6-0821 subm. to ICHEP 2004.
\bibitem{rapgap}
H.~Jung, \Journal{\CPC}{86}{1995}{147}.
\bibitem{h1fit2}
H1 Collaboration, \Journal{\ZPC}{76}{1997}{613}.
\bibitem{disjets}
H1 Collaboration, paper 6-0177 subm. to ICHEP 2004.
%\bibitem{kt} 
%S.~Ellis, D.~Soper, \Journal{\PRD}{48}{1993}{3160}; \\
%S.~Catani, Y.~Dokshitzer, M.~Seymour, B.~Webber,
\Journal{\NPB}{406}{1993}{187}.
\bibitem{disent}
S.~Catani, M.H.~Seymour, \Journal{\NPB}{485}{1997}{291}; erratum ibid. {\bf B510} (1997) 503.
%\bibitem{LUND} T.~Sj\"ostrand, \Journal{\CPC}{39}{1986}{347};\\
%               T.~Sj\"ostrand, M.~Bengtsson,
%               \Journal{\CPC}{43}{1987}{367}.
%\bibitem{PS}   M.~Bengtsson, T.~Sj\"ostrand,
%  \Journal{\ZPC}{37}{1988}{465}.

\bibitem{zeusdstar}
ZEUS Collaboration, \Journal{\PLB}{672}{2003}{3}.
\bibitem{dstar}
H1 Collaboration, paper 6-0178 subm. to ICHEP 2004.
\bibitem{hvqdis}
B.W.~Harris, J.~Smith, \Journal{\NPB}{452}{1995}{109};\\
L.~Alvero, J.C.~Collins, J.J.~Whitmore hep-ph/9806340.
\bibitem{zeusgp}
ZEUS Collaboration, paper 6-0249 subm. to ICHEP 2004.
\bibitem{frixione}  S.~Frixione, Z.~Kunszt and A.~Signer,
  \Journal{\NPB}{467}{1996}{399};\\
 S.~Frixione, \Journal{\NPB}{507}{1997}{295}.
\bibitem{kaidalov} A.~Kaidalov, V.~Khoze, A.~Martin, M.~Ryskin,
                   \Journal{\PLB}{567}{2003}{61}.
\end{thebibliography}
\end{document}